\input{aipcheck}

\documentclass[
    ,final            
  ]
  {aipproc}

\layoutstyle{8x11single}

\begin{document}

\title{On Bell's theorem, quantum communication, and entanglement detection}


\author{Marek \.Zukowski}{
  address={Institute of Theoretical Physics and Astrophysics, University
of Gda\'nsk, ul. Wita Stwosza 57, PL-80-952 Gda\'nsk, Poland}
}

\begin{abstract}
(A) Bell's theorem rests on a {\em conjunction} of three assumptions: realism, locality 
and ``free will''.  A discussion of these assumptions will be presented.
 It will be also shown that, if one adds to the assumptions  the
principle or rotational symmetry of physical laws, a stronger 
version of the theorem emerges.
(B) A link between Bell's theorem and communication complexity 
problems will be presented. This also includes experimental 
realizations, which surprisingly do not involve entanglement.
(C) A new sufficient and necessary criterion for entanglement of 
general (mixed) states will be presented. It is derived using the  same geometric
starting point as the inclusion of the symmetry in (A). The set of entanglement identifiers 
(EI's) emerging via this method contains entanglement witnesses (EW's), but they 
form only a subset of all EI's. Thus the method is more powerful than the 
one based on EW's.
\end{abstract}

\maketitle

\section{Introduction}
Quantum mechanics differs drastically in its mathematical formulation from any classical theory. One has an inherent randomness in the experimental observations. This is due to complementarity. If the horizontal polarization of a photon is  fully defined, this implies that the circular polarization measurements on such a photon would lead to totally random results. 
Entanglement reveals complementarity on a different level. If two qubits are maximally entangled, then the state of the full system is 
precisely defined, while the states of the individual qubits are totally undefined. 
It took a long time for the scientific community to accept, that entangled states are the essence of quantumness \cite{SCHROEDINGER}, and even more time to find direct applications of entanglement. 

Everything started in 1935 with the paradoxical paper of Einstein, Podolsky and Rosen (EPR) \cite{EPR}, followed by Bohr's response \cite{BOHR}. The debate soon faded away. Simply no one, including EPR, thought that the ideas concerning completion of quantum mechanics are in conflict with predictions of the theory. Thus the discussion was at this time only at a ``philosophical'' level. Arguments without observational consequences. The slogan ``shut up and calculate'' became a credo. The dormant subject re-emerged when Bell in 1964 showed that the EPR ideas are in a direct conflict with some experimentally testable quantum predictions \cite{BELL}. First observations of two-photon entanglement followed. Surprisingly, a quarter of century later, Greenberger, Horne and Zeilinger realized that three or more particle entanglements, despite intuitions based on the (vague) correspondence principle, lead to even more drastic violations of classicality  than two particle states \cite{GHZ}. A new experimental challenge emerged: to observe  multi-particle correlations, see e.g. \cite{ZHWZ}. The strange correlations of entanglement were found to 
have practical direct application in quantum cryptography, in the protocol of Ekert 1991 \cite{EKERT}. In 1993 the teleportation process was proposed \cite{TELEPORTATION}. A new multidisciplinary research discipline was born: quantum information. It aims at finding such quantum processes, which cannot have any classical model, and therefore any classical realization, and to harness these effects to obtain communication or computation devices with classically impossible properties.

Several topics concerning entanglement will be discussed here. 
\begin{itemize}
\item
The assumptions behind Bell's theorem \cite{GILL}, and over-interpretations of the implications of the theorem will be addressed. 
\item
Next, a fresh look at Bell's theorem will be presented, based on ref. \cite{ROTATIONAL}. It will be shown that introduction of additional grand principles, on which our understanding of physics is based, to the assumptions behind Bell's inequalities, leads to new inequalities. They are in some cases much more powerful than the standard ones. 
\item
Practical consequences of harnessing entanglement will be discussed using the example of reduction of communication complexity of some problems with distributed computation, \cite{COMPLEXITY-1}, \cite{COMPLEXITY-2}.   
\item
Bell inequalities are $100\%$ efficient in detecting entanglement only for pure states. Therefore we review  a new method of finding out entanglelement in mixed states \cite{ENTANGLEMENT}. It can be thought of as a non-linear extension of the entanglemnt witness method \cite{HORODECKI}. Some further generalizations will be presented.
\end{itemize}

\section{Bell's theorem}
In early sixties Bell conjectured, that if there is any conflict between quantum mechanics and the {\it realistic theories}\footnote{Realism, the very essence of classical physics, is  a view that any physical
system (i.e. also a subsystem of a compound system) carries full information, deterministic or (Kolmogorovian) probabilistic, on results of {\it all}
possible experiments that can be performed upon it.}, it may be confined to {\it local} local versions of such theories. {A theory is local if it assumes that information and influences cannot travel faster than light, i.e. it satisfies constraints of relativistic causality}. This led him to formulate his famous theorem, of profound scientific and philosophical consequences.

\subsection{Thought experiment}

We shall present a thought experiment that in the simplest form can describe an elementary experiment 
that one performs with entangled particles. The description will cover only observable features of the experiment.
  
At two  measuring stations $A$ and $B$, separated by a very long distance, one observes simultaneous flashy appearances of numbers $\pm1$ at the displays of controlling computers. They always appear in perfect coincidence, when observed in certain reference frame.
Right in the middle between the stations is something that we call ``source''. When it is absent, or not working, 
$\pm1$'s do not appear at the displays. The activation of the source always causes two $\pm1$ flashes, one at $A$, one at $B$.
They appear after a relativistic retardation time, or a bit later, with respect to the activation of the source, never before. 
The apparatuses at the detection stations  have a knob  which can be put in two positions: $m=1$ or $2$ at $A $ station, and 
 $n=1$ or $2$ at $B$. The procedure used to generate
the random choice of local knob positions equivalent to
\emph{independent, fair coin tosses}, thus: each of the four possible values of the pair $n,m$  are equally likely.
The coin tosses and knob settings are made at random times, and so often, so that the information on these is practically never available at the source during its activation periods (the tosses and settings cannot have a causal influence on the workings of source). When $\pm1$'s appear on the display, the local computer displays also the current setting. The data are stored, vary many runs of the experiment are performed.

\subsubsection{Assumptions of leading to Bell inequalities}

A minimalistic {\em local realistic} description of such an experiment would use the following assumptions \cite{GILL}:
\begin{itemize}
\item
\emph{Realism}. Any
mathematical-physical model which allows one to
use  \emph{eight} variables in the theoretical description of the experiments:
$A_{m,n}$, $B_{n,m}$, where $n,m=1,2$.  The variable $A_{m,n}$ gives the value, ${\pm 1}$, which could be obtained at station $A$, if the knob settings, at $A$ and $B$, were at positions $n,m$, respectively. Similarly, $B_{n,m}$ plays the same role for station $B$, under the same settings. This is equivalent to the assumption that a joint probability distribution of these variables\footnote{Note, that no hidden variables appear anywhere, beyond these. However, given a (possibly stochastic) hidden variables theory,
one will be able to define our eight variables as
(possibly random) functions of the variables in that theory.} is always allowed to exist: $p(A_{1,1},A_{1,2},A_{2,1},A_{2,2}; B_{1,1}, B_{1,2},B_{2,1} B_{2,2})$
\item
{\em Locality}. For all $n,m$:
\begin{equation}\label{e:locality}
A_{m,n}=A_{m},\quad B_{n,m}=B_n
\end{equation}
{\em The outcome which you would appear at A, under either setting, does not depend on which setting might be chosen at B, and vice versa}. Thus we are now assuming that $p(A_{1,1}, \ldots)$ reduces to $p(A_{1},A_{2},B_{1},B_{2})$
\item
  \emph{Freedom, or ``free will''},
often only a tacit assumption,
\begin{equation}\label{e:freedom}
(n,m)~~\mbox{is statistically independent of}~~(A_{1},A_{2},B_{1},B_{2})\qquad
\mbox{(freedom).}
\end{equation}
That is, the probability distributions satisfy 
\begin{equation}
p(n,m, A_{1},A_{2},B_{1},B_{2})=P(n,m)p(A_{1},A_{2},B_{1},B_{2}) \label{IDEPENDENCE}
\end{equation}
The choice of settings in the
two randomizes,  $A$ and $B$,
is causally separated from the locally realistic mechanism, which produces
the potential outcomes.
\end{itemize}

Contained in the above is an assumption of \emph{local disturbances}. When setting labels $m$, $n$ are sent to the measurement devices,
they will likely cause some further unintended disturbance: {\em any disturbance at A, as far as it influences the outcome at A,
is not related to the coin toss nor to the potential outcomes at B, and vice versa}.

\subsubsection{Lemma: Bell inequality}
The probabilities, $\Pr$, of the four propositions, $A_n=B_m$, satisfy 
\begin{equation}\label{e:Edelta}
\mathrm \Pr\{A_1=B_2\}
-\Pr\{A_1=B_1\}-\Pr\{A_2=B_1\}-\Pr\{A_2=B_2\}~\le~0.
\end{equation}
Proof: only four, or two, or none of the propositions, in the left hand side of the inequality can be true, thus (\ref{e:Edelta}).

If the observation settings are totally random (dictated by ``coin tosses''), $P(n,m)=\frac{1}{4}$.
According to all our assumptions  $P(A_n=B_m\mid n,m)=P(n,m)\Pr\{A_n=B_n\}=\frac{1}{4}\Pr\{A_n=B_m\}$. 
Therefore we have a  Bell inequality: under the conjunction of the assumptions for the experimentally accessible probabilities one has
\begin{equation}\label{e:bellinequ}
B=P(A_1=B_2\mid 1,2)-P(A_1=B_1\mid 1,1)
-P(A_2=B_1\mid 2,1)-P(A_2=B_2\mid 2,2)~\le~0.
\end{equation}

\subsubsection{The theorem}
Quantum mechanics predicts for some experiments satisfying all the features of the thought experiment   $B=\sqrt 2 -1,$ which is $\gg 0$.
{\em Hence, Bell's theorem:
if quantum mechanics holds, local realism, defined by the {\em full set} of the above assumptions, is untenable.}
What is perhaps even more important, it seems that we are approaching the moment, in which one could have as perfect as possible laboratory realization of the thought experiment (locality loophole was closed in \cite{WEIHS}, whereas detection loophole in \cite{XXX}). Hence local realistic approach to description of physical phenomena is close to be shown untenable too. 

\subsubsection{Philosophy or physics?}

The assumptions behind Bell inequalities are often criticized as being ``philosophical''. However, the whole (relativistic) classical physics is realistic (and local). Thus we have an important exemplary realization of the postulates of local realism in form of grand physical theories. Further, one could define philosophical propositions as those which {\em are not} observationally or experimentally falsifiable (at the given moment of the development of human knowledge), or in pure mathematical theory are not logically derivable. Therefore, the {\em conjunction } of all assumptions of Bell inequalities is not a philosophical statement, as it is testable both experimentally and logically (within, known at the moment, mathematical formulation of fundamental laws of physics).
Thus, Bell's theorem  removed the question of possibility of local realistic description from the realm of philosophy. Now this is just a question of a good experiment. The atomic hypothesis was a philosophical proposition for centuries, now it is not anymore. Of course, if one defines the realm of philosophy as the set of subjects discussed by philosophers, then even the $C^*$ algebraic quantum mechanics is a part of it, not only local realism. 

The other criticism is formulated in the following way. Bell inequalities can be derived using a single assumption of existence of joint probability distribution for the observables involved in them, or that the problablity calculus of the experimental propositions involved in the inequalities is of Kolmogorovian nature, and nothing more. But if we want to apply these assumptions to the thought experiment we stumble on the following question: {\em does the joint probability take into account full experimental context or not?}. The experimental context is in our case (at least) the full state of the settings $(m,n)$. Thus if we use the same notation as above for the realistic values, this time applied to the possible results of measurements of observables, initially we can assume existence of only $p(A_{1,1},A_{1,2},A_{2,1},A_{2,2}; B_{1,1}, B_{1,2},B_{2,1} B_{2,2})$. 

Let us discuss this from the quantum mechanical point of view, only because such considerations have a nice formal description within this theory, familiar to all physicists. Two observables, say $\hat{A}_1\otimes\hat{B}_1$ and $\hat{A}_1\otimes{B}_2$, as well as other possible pairs are functions of two different {\em maximal} observables for the whole system (which are non-degenerate by definition). If one denotes such a maximal observable linked with $\hat{A}_m\otimes\hat{B}_n$ by $\hat{M}_{m,n}$ and its eigenvalues by $M_{m,n}$ the existence of the aforementioned joint probability is equivalent to the existence of a $p(M_{1,1},M_{1,2},M_{2,1},M_{2,2})$ in form of a proper probability distribution. Only if one assumes additionally context independence, this can be reduced to the question of existence of $P(A_{1},A_{2},B_{1},B_{2})$, where $A_{m}$ and $B_{n}$ are eigenvalues of $\hat{A}_m\otimes\hat{I}$ and $\hat{I}\otimes\hat{B}_n$, where it turn $\hat I$ is the unit operator for the given subsystem. While context independence is physically doubtful, when the measurements are not spatially separated, and thus one can have mutual causal dependence, it is well justified for spatially separated measurements. I.e., {\em locality} enters our reasoning, whether we like it or not. Of course one cannot derive any Bell inequality of the usual type if the random choice of settings is not independent of the distribution of  $A_{1},A_{2},B_{1},B_{2}$, that is without (\ref{IDEPENDENCE}).

\subsubsection{The assumptions as a communication complexity problem}

For those for whom even these arguments smell of philosophy one can formulate  the Bell theorem in form of a technical problem in computer science.

Assume that we heave two partners $P_k$, with $k=1,2$. They share certain joint classical information strings of arbitrary lengths and/or some programs, or protocols of action. All these will be collectively denoted as  $\lambda$.  But, no communication between them is allowed. Each gets a one bit random number $x_k$, known only to him/her (e.g., they generate them by a ``coin toss'', the process must be stochastically independent of anything else in the problem). The {\em individual} task of each of them is to produce  a one bit number $I_k(x_k, \lambda)$, and communicate only this one bit to a Referee who just compares the received bits. There is no restriction on the form and complication of the functions $I_k$, or any actions taken to define the values, but any communication between the partners is absolutely not allowed.
The {\em joint} task of the partners is to find a way which under, the constraints listed above, and without any cheating, allows to have after very many repetitions of the procedures (each starting with establishing a new shared $\lambda$) the following functional dependence of the  probability that their bits sent to the Referee are equal:
\begin{equation}
P_Q\{I_1(x_1)=I_2(x_2)\}=\frac{1}{2}+ \frac{1}{2}\cos\big[-\pi/4+(\pi/2)(x_1+x_2)\big].
\end{equation}
But this is impossible with the classical means at their disposal, and without communication. Simply
because with their protocol
\begin{equation}\label{e:bellinequ}
B=\Pr\{I_1(1)=I_2(1)\}-\Pr\{I_1(0)=I_2(0)\}-\Pr\{I_1(1)=I_2(0)\}-\Pr\{I_1(0)=I_2(1)\}
~\le~0.
\end{equation}
whereas the  value for $P_Q$ is $\sqrt{2}-1$. With the possibility of sending during the communication stages (when $\lambda$ is collected) to each of the partners a qubit from a maximally entangled pair, one can obtain on average $P_Q$. The messages  send to the Referee encode in bit encoding the local result of measurement of Pauli observables $\vec{n}\cdot\vec{\sigma}$, where $||\vec{n}||=1$, and the local measurement directions are suitably chosen as functions of $x_1$ and $x_2$.   

\subsection{Consequences of Bell's theorem}

Bell's inequalities, when violated by quantum predictions give a clear cut indication of entanglement, as any pure entangled state violates a certain Bell inequality \cite{GISIN}.      
Such violations  imply that the underlying {\em conjunction of assumptions of realism, locality and ``free will''}  is not valid, 
and {\em nothing more}.  

It is often said that the violations indicate ``(quantum) non-locality''. However
if one  wants {\em non-locality} to be {\em the} implication, one has to assume ``free will'' and realism. But this is only at this moment a philosophical choice (there is no way to falsify it). {\em It is not a necessary condition for violations of Bell inequalities.} 

Still, in such a case the implication should be: {\em classical (realistic) non-locality}.
However, due to complementarity quantum formalism does not use realism\footnote{Recently a certain class of ``reasonable'' non-local realistic theories was ruled out experimentally (under fair sampling assumption), \cite{NATURE}. }. 
Due to many reasons quantum predictions for entangled states do not allow instantaneous signaling (this is surprising, that this so even in the standard non-relativistic formulation  of the description of quantum states!). So why quantum-non-locality?

\subsection{Generalized Bell inequalities}

In this section we shall show that if one introduces one more, quite a natural assumption, except from those which were discussed above,
one gets a new version of Bell's inequalities, and hence of Bell's theorem, which in case of some states reveals much more non-classicality than the standard formulation.   

This assumption could be called  -  
``rotational invariance of physical laws''. As we know leads to conservation of total angular momentum of isolated systems. 
We shall put it in the following way. Take a correlation experiment on $N$ spatially separated spins $1/2$. 
The $N$ particle correlation function is defined as the following average
$
E(\vec n_1, \vec n_2, ..., \vec n_N) = \langle \prod_{k=1}^nr_k(\vec n_k) \rangle_{avg},
$
where $r_j(\vec n_j)$ - local result of a dichotomic measurement, equal to $\pm 1$, and $\vec n_j$ are the local measurement directions - imagine a Stern-Gerlach type experiment. Our ``rotational invariance'' assumption allows such functions to have only the following explicit scalar form 
\begin{equation}
E(...)= \hat T \cdot (\vec n_1 \otimes \vec n_2 \otimes ... \otimes \vec n_N).\label{ROTATIONAL}
\end{equation}
One could imagine more complicated forms, but here $E$ is assumed to be linearly dependent on $\vec n_i$'s, just like it is quantum mechanics. 
The components of correlation tensor, $\hat T $,  are given by
$
T_{i_1...i_N}=E(\vec x_{1}^{(i_1)},\vec x_{2}^{(i_2)}, ..., \vec x_{N}^{(i_N)}),
$
where $i_k=1,2,3$ and $\vec x_{k}^{(i_k)}$ are three arbitrary orthogonal directions for $k$-th observer.

The generalized Bell's inequality derived in \cite{ROTATIONAL} reads as follows: 
for local realistic rotationally invariant correlation functions one has
\begin{equation}
S=\sum_{i_1,i_2,\ldots,i_N=1,2}T_{i_1i_2...i_N}^2\leq \big({4}/{\pi }\big)^N E_{max},
\label{EEvalue}
\end{equation}
where $E_{max}$ is the maximal value of the correlation function, if one restricts the measurements to directions spanned out by those 
{pairs} used in the left hand side (that is,  observer  $k$ restricts measurement directions to those spanned by $\vec{x}_k^{(i_k)}$, with $i_k=1,2$). 

The inequality is derived by estimating the upper bound for a scalar product of
\begin{eqnarray}
&E_{LR}(\vec{n}_1,\vec{n}_2,\ldots,\vec{n}_N)&\nonumber\\
&=\int d\lambda \rho(\lambda)
I^{(1)}(\vec{n}_1,\lambda)I^{(2)}(\vec{n}_2,\lambda)\cdots
I^{(N)}(\vec{n}_N,\lambda),&\label{LHVcofun}
\nonumber\end{eqnarray}
where
$I^{(j)}(\vec{n}_j,\lambda)=\pm1$,
with
\begin{eqnarray}
&E(\alpha_1,\alpha_2,\ldots,\alpha_N)=
\hat{T} \cdot \vec{n}_1(\alpha_1)
\otimes
\vec{n}_2(\alpha_2)
\otimes\cdots
\otimes\vec{n}_N(\alpha_N).&
\label{INVARIANT}
\end{eqnarray}
The correlation function $E_{LR}$ has a structure which is allowed for local hidden variable theories, which can always model any local realistic theory. The hidden variable $\lambda$ can be of any form, thus the integration symbol stands for integration and/or summation over as many variables as one wishes.
For the estimate one uses the following parametrization of the measurement directions.  
$
\vec{n}_j(\alpha_j)=\cos \alpha_j \vec{x}_j^{(1)}
+\sin \alpha_j \vec{x}_j^{(2)}.$
The scalar product reads
\begin{eqnarray}
& (E_{LR}, E) =\int_0^{2\pi}d\alpha_1
\int_0^{2\pi}d\alpha_2\cdots
\int_0^{2\pi}d\alpha_N E_{LR}(\alpha_1,\alpha_2,\ldots,\alpha_N)
E(\alpha_1,\alpha_2,\ldots,\alpha_N).&\nonumber\\
\label{TE}
\nonumber\end{eqnarray} 

The geometric intuition behind all this is the following one \cite{ZUKOWSKI-1993}. The aim is compare two correlation functions. 
One has the structure required by local realism, $E_{LR}$, the other one, $E$,  the structure required by rotational invariance, given by (\ref{ROTATIONAL}). Note that such a form have the quantum predictions, $E_{QM}$.
If one defines a scalar product and one has 
\begin{equation}
(E_{LR},E_{QM})\leq B< (E_{QM},E_{QM}), \label{IDEA}
\end{equation}
then obviously  $E_{QM}\neq E_{LR}$. That is, in such a case 
the correlation function cannot be reproduced by a local realistic model.

For example in the case of a noisy GHZ state, given by
 $V |\psi_{GHZ} \rangle \langle
\psi_{GHZ}|+(1-V) \rho_{noise},$ 
where 
$\rho_{noise} = \frac{1}{2^N} \hat{\bf 1},$ 
one has  $T_{max}=V,$ and  
$\sum_{i_1,i_2,\ldots,i_N=1,2}T_{i_1i_2...i_N}^2=V^2 2^{N-1}.$
Thus, $(E_{LR}, E) \leq 4^N V$ whereas
$(E,E) = \pi^N V^2 2^{N-1}$.
Thus local realism and rotational invariance 
 principle exclude local realistic models for $V > 2(\frac{2}{\pi})^N$. 
In the case of standard Bell inequalities one must have $V \geq \frac{1}{\sqrt{2^{N-1}}}$ to violate them. 
The new thresholds,   with respect to the number of particles, are thus exponentially more restrictive than the standard ones.

Note that,  locality is a direct consequence of the Lorentz transformations (boosts). A subgroup of the full Poincar{\'e} group, rotations, introduces an additional constraint on local realistic models. 
This constraint is introduced here on the level of correlation functions, i.e. {\em after averaging over hidden variables}, ($\lambda$'s). In contradistinction the {\em locality condition is introduced for every value of} $\lambda$. 
Note that, if one assumes {\em locality after averaging over hidden variable theories}, one gets the so called {\em ``no-signaling condition''}. Such a condition works well with realistic models, as there exist realistic models of quantum mechanics which are non-local on the hidden variable level.  Had we worked with equivalently soft formulation of locality (on the level of averages) we would not have been able to exclude any form of realistic theories. Thus, as such a soft form of imposing symmetries of laws of physics has so drastic consequences, 
 one could ask  
{\em which other symmetries further constrain  local realistic theories?}

\section{Quantum reduction of communication complexity}
Entanglement violating a Bell inequality can always be used to find a better-than-any-classical  solution to some problems requiring communication between separated partners.

Reduction of communication complexity is a standard problem in classical informatics (Yao, 1979, \cite{YAO}), with obvious applications in communicational and computational networks.   What will be shown below is a development and generalization of the ideas of \cite{BURN} and \cite{GALVAO}, presented in \cite{COMPLEXITY-1} and \cite{COMPLEXITY-2}.

Imagine several spatially separated partners, $P_1$ to $P_N$. Each of whom has some data  known to him/her only, denoted here as $X_i$, with $i=1,...,N$. They face a joint task: to compute the value of a function $T(X_1,...,X_N)$. This function depends on all data. Obviously they can get the value of $T$ by sending all their data to partner $P_N$, who does the calculation and announces the result. But are there ways to reduce the amount of communicated bits (communication complexity of the problem)? For very many tasks this is so. But there are tasks that can be solved, under say communication restriction to $N-1$ bits, only if the th e protocol utilizes quantum laws. Such an example will be presented below.

\subsubsection{Example}

Assume that partner $P_k$ has a two bit string $X_k=(z_k,x_k)$.  We shall consider specific task functions which have the following form 
$$T=f(x_1,...,x_N)(-1)^{\sum z}, $$ where $\sum z$ denotes $\sum_{k=1}^N z_k$. The partners know also the probability distribution (``promise'') of 
  the bit strings (``inputs''). We shall consider only distributions, which are completely random with respect to $z_k$'s, that is a class of  the form 
$p(X_1,...,X_N)=2^{-N}p'(x_1,...,x_N)$.
Communication is restricted to $N-1$ bits. Assume that we ask the last partner to give the answer $A$ to the question what is the value of $T$. 

For simplicity, we shall introduce now $y_k=(-1)^{z_k}$. We shall use $y_k$ as a synonym of $z_k$.
Since $T$ is proportional to $\prod_k y_k$, the final answer $A$, equal to $\pm 1$,  is completely random if it does not depend on every $y_k$. 
Thus,  information on $z$'s from all $N-1$ partners must somehow reach $P_N$. 
Therefore the only communication ``trees'' which might lead to a success are those in which each $P_k$ sends only 
a one-bit message $m_k$. Again we introduce: $e_k=(-1)^{m_k}$, and will treat is as synonym of $m_k$ .  

The average success of a communication protocol can be measured with the following fidelity function 
\begin{equation}
F=\sum_{X_1,..., X_N}p(X_1,...X_N)T(X_1,...X_N)A(X_1,...X_N),\end{equation}  or equivalently
\begin{eqnarray}
&F=\frac{1}{2^N}\sum\limits _{x_1, \ldots, x_N=0,1}p'(x_1, \ldots,
x_N)f(x_1, \ldots, x_N)& \nonumber\\ &\times \sum\limits _{y_1, \ldots, y_N=\pm1}\prod_{k=1}^N{y_k} A(x_1, \ldots, x_N;y_1, \ldots, y_N ).& \label{FIDELITY}
\end{eqnarray}
The probability of success is $P =(1+F)/2$.

 The first steps of a derivation of the reduced form of the fidelity function for an optimal protocol will now be presented (the reader may reconstruct the other steps or consult reference \cite{COMPLEXITY-2}).
In a classical protocol the answer $A$ of the partner $P_N$ can depend
on the local input
$y_N$, $x_N$, and  messages, $e_{i_1},..., e_{i_l},$ received {\em directly}
from  partners $P_{i_1}, ...P_{i_l}$: 
 \begin{equation}A=A(x_N,y_N,
{e}_{i_1},..., {e}_{i_l}).\end{equation}
Let us fix $x_N$, and treat $A$ as a function $A_{x_N}$ of the remaining ${l+1}$  dichotomic
variables $$y_N,{e}_{i_1},..., {e}_{i_l}.$$ That is, we treat now $x_N $ as a fixed index.
 All such functions can be thought of as $2^{l+1}$ dimensional vectors, because the values of each such a function form a sequence of the length equal to the number of elements in the domain. In the $2^{l+1}$ dimensional space containing  such functions
 one has {\em an orthogonal basis}
given by \begin{equation}V_{j,j_1,...j_l}(y_N,
{e}_{i_1},..., {e}_{i_l})={y_Nj}\prod_{k=1}^l {{e}_{i_k}{j_k}},\label{QQQ}\end{equation}
where $j,j_1,...j_l=0,1$. 
Thus, one can expand $A_{x_N}$ with respect to this basis
and the expansion coefficients read
\begin{equation}
c_{jj_1,...j_l}(x_N)=\frac{1}{2^{l+1}}
\sum_{y_N,
{e}_{i_1},..., {e}_{i_l}=\pm1}{e}_{x_N}{V}_{j,j_1,...j_l}.\end{equation}
Since $|A_{x_N}|=|V_{j,j_1,...j_l}|=1,$ one has $|c_{jj_1..j_l}(x_N)|\leq 1.$
We put the expansion to $F$ and get 
\begin{eqnarray}
&F=\frac{1}{2^N}\sum\limits _{x_1, \ldots, x_N=0,1}g(x_1, \ldots, x_N)
 \sum\limits _{y_1, \ldots, y_N=\pm1}\prod_{h=1}^Ny_h 
 &\nonumber\\
&\times
 \big[\sum_{j,j_1,...j_l=0,1}c_{jj_1,...j_l}
(x_N)y_N^j \prod_{k=1}^l
{e}^{j_k}_{i_k} \big],&\nonumber \\ 
\end{eqnarray}
where $g(x_1, \ldots, x_N) \equiv f(x_1, \ldots, x_N)p'(x_1, \ldots, x_N).$
As, $\sum_{y_N=\pm1}y_Ny_N^0=0$, and  $\sum_{y_{k}=\pm1}y_{k}{e}_{k}^0=0$, only the  term 
  with $j,j_1,...j_l=1$  can give a non-zero contribution
to $F$.
 Thus,  $A$ in $F$ can be replaced by
 \begin{equation} A'=y_Nc_N(x_N)\prod_{k=1}^l {e}_{i_k}, \label{XXX1}\end{equation}
 where  
$c_N(x_N)$ stands for $c_{11...1}(x_N)$. 
Next,  notice that, e.g.,  ${e}_{i_1}$,
can depend only on $x_{i_1}$, $y_{i_1}$ (local data) and the messages obtained by $P_{i_1}$ from a subset of partners: $e_{p_1},..., e_{p_m}.$ This set does not contain any of the $e_{i_k}$'s of the formula (\ref{XXX1}) above.
In analogy with $A$, the function ${e}_{i_1}$, for a fixed $x_{i_1}$, can be treated as a vector, and thus can be
expanded in terms of  orthogonal basis functions (of a similar nature as (\ref{QQQ})), etc.
Again, the expansion coefficients satisfy $|c'_{jj_1,...j_m}(x_{i_1})|\leq1$.
If one puts this into $A'$, one obtains a new form of $F$, which effectively depends on $A''=c_N(x_N)c_{i_1}(x_i)\prod_{k=2}^l {e}_{i_k},$ where 
 $c_{i_1}(x_i)$ stands for $c'_{11...1}(x_{i_1})$, and its modulus is again bounded by $1$. Note that, $y_N$ and $y_{i_1}$ disappear, as $y_k^2=1$. 

As each message appears in the product only once, we continue this 
procedure of expanding those messages which depend on earlier messages, till it halts. 
The final reduced form of the formula for the fidelity of an optimal protocol reads
\begin{equation}
F=\mathop{\sum}\limits _{x_1, \ldots, x_N}g(x_1, \ldots,
x_N)\prod_{n=1}^{N}c_n(x_n),
\label{BELL}
\end{equation}
 with $|c_n(x_n)|\leq 1$.
 Since $F$ in eq. (\ref{BELL}) is linear in every
 $c_n(x_n)$, its extrema are at the limiting values  $c_n(x_n)=\pm1$.
  {\em In other words, a Bell-like inequality $|F| \leq \textrm{Max}({F})\equiv B(N)$ gives the
upper fidelity bound}\footnote{Note, that the above derivation shows that optimal classical protocols include one in which partners $P_1$ to $P_{N-1}$
 send to 
$P_N$ one bit messages which encode the value of $e_k=y_kc(x_k)$, where $k=1,2,...N-1$.}.

\subsection{Quantum solutions}
The  inequality for $F$ suggests that some problems may have
quantum solutions, which surpass any classical ones in their fidelity.
Simply one may use an entangled state $|\psi\rangle$ of $N$ qubits. Send to each of the partners one of the qubits.
In a protocol run all $N$ partners make measurements on the local qubits, the settings of which are determined by $x_k$\footnote{They measure a certain qubit observable 
$\vec{n}_k(x_k)\cdot\vec\sigma$.}. The measurement results $\gamma_k=\pm1$ are multiplied by $y_k$, and the partner $P_k$, for $1\leq k \leq N-1$ send a  bit message to $P_N$ encoding the value of $m_k=y_k\gamma_k$. The last partner calculates $y_N\gamma_N\prod_{k=1}^{N-1}m_k$, and announces  this as $A$. The average fidelity of such a process is
\begin{eqnarray}
& F=\mathop{\sum}\limits _{x_1, \ldots, x_N}g(x_1, \ldots,
x_N)
\langle\psi|\otimes_{n=1}^{N}(\vec{n}_k(x_k)\cdot\vec{\sigma}_k) |\psi\rangle,\nonumber\\
\label{BELL-VIOLATED}
\end{eqnarray}
and in some cases equals even $1$.

For some tasks the quantum vs. classical fidelity ratio grows exponentially with  $N$. 
This is so e.g. for the so-called {\em modulo-4 sum} problem. Each partner receives a two-bit input
string  $\left(X_k=0,1,2,3; k = 1,\ldots,N \right).$ The $X_k$'s are distributed so that
$(\sum_{k=1}^{N} X_k){\textrm{mod} 2} = 0.$
 The task is\footnote{It can be formulated in terms of a task function
$T = 1-(\sum_{k=1}^{N} X_k)\textrm{mod} 4.$  An alternative formulation of the problem reads
 $f =\cos(\frac{\pi}{2}\sum_{k=1}^{N}x_k)$ with
$p'  = 2^{-N+1}|\cos(\frac{\pi}{2}\sum_{k=1}^{N}x_k)|.$}: $P_N$ must tell whether the sum modulo-4 of all inputs is  0
or 2.

 For this problem  the classical fidelity bounds decrease exponentially with $N$, that is $B(F)\leq2^{-K+1},$ where $K=N/2$ and $K=(N+1)/2$ for even and odd number of parties, respectively.   
If one uses the $N$ qubit GHZ states $\frac{1}{\sqrt{2}}(|0,...,0\rangle+|1,...,1\rangle)$, where $\langle0|1\rangle=0$, 
and suitable pairs of local settings, the associated Bell inequality can be violated maximally. Thus,  one has a quantum protocol which always gives the correct answer.

Surprisingly, one can also show a version of a quantum protocol without  entanglement \cite{GALVAO}. The partners exchange a single qubit, $P_k$ to $P_{k+1}$ and so on, and each of them makes a suitable unitary transformation on it (which depends on $z_k$ and $x_k$). The partner $P_N$, who receives the qubit as the last one, additionally performs a dichotomic measurement. The result he/she gets is equal to $T$. For details see, including an experimental realization see \cite{COMPLEXITY-2}. The obvious conceptual advantage of such a procedure is that the partners exchange a single qubit, which can carry at most one bit of readable information. In contrast with the protocol involving entanglement, no classical transfer of any information is required, except from the announcement by $P_N$ of his measurement result!

In summary, if one has a pure entangled state of many qubits (this can be generalized to more complicated systems), then there exist a Bell inequality which is violated by this state. This  inequality has some coefficients $g(x_1,...,x_n)$, which can always be renormalized in such a way that $$\sum_{x_1,...,x_n}|g(x_1,...,x_n)|=1.$$ The function $g$ can always be interpreted as a dichotomic function $f(x_1,...,x_n)=\pm 1$
times a probability distribution $p'(x_1,...,x_n)=|g(x_1,...,x_n)|$. Thus we can construct a communication complexity problem of the type discussed above, with task function $T=\prod_i^N y_i f$. All this can be extended beyond qubits, see \cite{PATEREK}.

\section{Geometrical separability vs. entanglement criteria}
 
The simple idea \cite{ZUKOWSKI-1993}, which is a root of (\ref{IDEA}), that for two vectors, in an arbitrary space with a scalar product, say $\vec{e}$ and $\vec{s}$,
if  one has 
$\vec{s}\cdot\vec{e}<\vec{e}\cdot\vec{e}=||\vec{e}||^2$
then
$\vec{e}\neq\vec{s}$, can be also used as a starting point for a derivation of a necessary and sufficient criterion for a general quantum state to be separable. Such criteria have to replace, for mixed states, Bell inequalities, because there exist mixed entangled states which do have local realistic models \cite{WERNER}.

Let us constrain our discussion only to multi-qubit systems.
Any quantum state
can be decomposed in the following way
\begin{equation}
\rho = \frac{1}{2^N} \sum_{j_1,...,j_N=0}^3 T_{j_1...j_N}
\sigma_{j_1} \otimes ... \otimes \sigma_{j_N}, \label{STATE}
\end{equation}
where $\sigma_{j_n}$ is the
$j_n$-th local Pauli operator of the $n$-th party, for $j_n=1,2,3$, and $\sigma_0=I$. The real expansion coefficients $T_{j_1...j_N}$ form
an object which will be called a ``generalized correlation tensor''.
A state $\rho$ is separable if it can be decomposed as:
\begin{equation}
\rho_{\mathrm{sep}} = \sum_{i} p_i \rho_i^{(1)} \otimes ... \otimes
\rho_i^{(N)},
\end{equation}
with $p_i \ge 0$ for all $i$, and $\sum_i p_i = 1$.
Thus every fully separable state
is specified by a  generalized tensor $\hat T^{\mathrm{sep}} =
\sum_{i} p_i \hat T^{\mathrm{prod}}_i$,
where $\hat T^{\mathrm{prod}}_i = \hat T^{(1)}_i \otimes ... \otimes \hat
T^{(N)}_i$
and each $\hat T^{(k)}_i$ describes a pure qubit state.

One can introduce a certain generalized scalar product of the correlation tensors which utilizes only the $N$ particle components of $T$: for tensors $S$ and $Q$ one can e.g. define
$(\hat S,\hat Q) =  \sum_{j_1,...,j_N=1,2,3} S_{j_1...j_N}Q_{j_1...j_N}$. That is, in this scalar product enter the only the components of $\hat{T}$ of the previous section, those that pertain only to averages of tensor products of {\em proper} Pauli matrices.
For any separable tensor $\hat{T}$, one has
\begin{equation}
 {(\hat T,\hat T)=||\hat T||^2}\leq{T^{\max}},
\label{SIMPLE_E}
\end{equation}
where 
\begin{equation}
T^{\max}=
 \max_{\hat T^{(1)} \otimes... \otimes \hat T^{(N)}} 
\sum_{j_1,...,j_N=1,2,3} T_{j_1...j_N} T_{j_1}^{(1)} \cdot ... \cdot T_{j_N}^{(N)},
\label{MAXSC}
\end{equation}
One can derive this condition using the geometrical idea opening this section and properties of convex combinations in scalar products.  Note that $T^{max}$ is the highest possible value of a component of $\hat{T}$ (general components of $\hat{T}$ are given by $(\hat{T},\hat n^{(1)} \otimes... \otimes \hat n^{(N)} )$, where $\hat{n}$'s represent unit three dimensional vectors), and that $\hat{T}^{(i)}$'s used here are effectively normalized three dimensional vectors.  

That is,  we have a simple necessary condition for separability in form of (\ref{SIMPLE_E}). 
It is quite powerful. If violated, it can detect entanglement, e.g. of {\em all} ``Werner'' two qubit states (mixtures of pure noise with a maximally entangled states). No linear entanglement witness has this property. The condition constitutes a kind of shell which has all separable states inside or on the surface, and entangled Werner states outside. Whereas, an entanglement witness defines a hyperplane with {\em all} separable states on one side, and {\em some} entangled states on the other.

To get an even stronger condition, one can use all possible generalized scalar products, which are defined with respect to 
some generalized metric tensor $G$ (which does not have to be strictly positively defined).
If
$|(s, Ge)|<(e,Ge)=||{e}||_G^2,$
for a $G\geq 0$
then
${e}\neq{s}$.
With this insight we get (almost) immediately:
for every entangled state, $\rho_{ent}$, one can find a generalized scalar product,  defined by a non-negative superoperator $G$, such that
\begin{equation} Max_{\rho_{sep}}|Tr\rho_{sep}G\rho_{ent}|<Tr\rho_{ent}G\rho_{ent},\end{equation}
where the maximum is over pure separable states.
The intuitive ground for this statement can be put as follows.  The separable states form a compact convex subset of the space of self-adjoint operators \footnote{The space of density operators is a subset of the full space of self-adjoint operators.
 Any self-adjoint operator is a linear combination of a basis set of self adjoint operators. This is due to the fact that 
 in the space of operators one can define the following ``Hilbert-Schmidt" scalar product 
 $
 (A,B)=TrA^{\dagger}B.
 $
 For two self-adjoint operators such a scalar product is always real.
 Every operator can be expressed in the following way:
 $
 A=\sum_{k,l}A_{kl}|k\rangle\langle l|.
 $
 One can build the following $N^2$ self-adjoint operators:
 $\frac{1}{2}(N^2-N)+N=\frac{1}{2}N{N+1}$ operators of the form
 $
 B(kl)=|k\rangle\langle l|+|l\rangle\langle k|,
 $
 with $k\geq l$
 and $\frac{1}{2}(N^2-N)$ operators
 $
 B(kl)=i|k\rangle\langle l|-i|l\rangle\langle k|,
$
 for $k<l$. It is elementary to show that  they form a basis in the space of operators. Also for any self adjoint $A$ all numbers $(A,B(kl) )=Tr(A^\dagger B(kl))$ are real. That is the the space of self-adjoint operators is isomorphic with a $N^2$
 dimensional space of {\em real} vectors, since just like those it is one-to-one representable by sequences of real numbers containing $N^2$ elements. Thus we can use all our intuitions, which are associated with real vector spaces, to the space of self-adjoint operators.}.
 Therefore one can always define a generalized coordinate system such that it has its origin within this separable subset, e.g. at $\rho_{noise}$. For every entangled state there must exist a generalized coordinate, which is of a larger modulus  than the same coordinate of any separable state. The generalized metric operators simply give, if necessary, an excessive weight to this coordinate, so big that the above inequality holds. In the case of entangled states most  stubbornly  resisting detection by this criterion the $G$ metric can be non-zero only for this single coordinate\footnote{Such G superoperators are {\em effectively} equivalent to an entanglement witness for the state}. For a more formal proof see Badziag el al. \cite{ENTANGLEMENT}. 

This criterion defines a new type on entanglement identifiers in form of metric superoperators $G$. 
The set of entanglement identifiers defined by these
conditions is strictly richer than the set of entanglement witnesses.
The examples given in \cite{ENTANGLEMENT} indicate that
many identifiers not corresponding to any standard entanglement witness
are particularly interesting. The condition can be tailored to all possible types of separability problems, and all (finite) dimensions of the 
subsystems.

\section{summary}
As it was shown above entanglement leads to processes which are of a non-classical character. This manifests itself by violations of 
Bell inequalities. However, one must be very careful in interpreting this fact. 

It was also shown that if one has entanglement which violates a Bell inequality, then one can devise a communication complexity problem for which the Bell  like inequality gives the fidelity bound for all possible classical protocols, and that  there exists a quantum communication protocol which under identical communicational restrictions violates the classical bound. In short a Bell inequality violating entanglement is resource which can be always applied to some potentially useful tasks.

 However, in some situations entanglement cannot be detected via Bell inequalities, but still one can distill it, if one has many copies of equivalently prepared systems \cite{DISTILL}, to get less copies with stronger entanglement, which do violate Bell inequalities. To detect such a hidden entanglement one must resort to different methods. An example of such a method was given above.   

\begin{theacknowledgments}
This work is a part of the EU 6-th Framework programmes QAP and SCALA.
It has been done at {\em National Quantum Information Centre of Gdansk}. 
  
\end{theacknowledgments}




\begin{thebibliography}{9}
\bibitem{SCHROEDINGER}
 E. Schroedinger, Naturwissenschaften {\bf 23}, 807-815; {\em ibidem} {\bf 23}, 823-831; {\em ibidem}  {\bf 23}, 844-852 (1935). 

\bibitem{EPR}
A.~Einstein, B.~Podolsky, and N.~Rosen: Phys. Rev.
   {\bf 47}, 777-784 (1935).
 

\bibitem{BOHR}
N. Bohr,  {Phys. Rev.} {\bf 48}, 696-705 (1935).

 
\bibitem{BELL}
J.S. Bell, Physics {\bf 1}, 195-201 (1964);
J. Clauser, M. Horne, A. Shimony, R. Holt, Phys. Rev. Lett. {\bf 23}, 880-883 (1969).
 

\bibitem{GHZ}
D. M. Greenberger, M. A. Horne, and A. Zeilinger,
in \emph{Bell's Theorem, Quantum Theory and Conceptions of the Universe},
edited by M. Kafatos, Kluwer Academic, Dordrecht, 1989.

\bibitem{ZHWZ}
M. \.Zukowski, A. Zeilinger and H. Weinfurter, Ann. N. Y. Acad. Sci. {\bf 91},  755-763 (1995);
M. \.Zukowski, A. Zeilinger, M. A. Horne, A.K. Ekert, "Event Ready Detectors" Bell Experiment via Entanglement Swapping, Phys. Rev. Lett. {\bf 71} 4297-4300 (1993).

\bibitem{EKERT}
 A.K.~Ekert: Phys. Rev. Lett. \textbf{67}, 661--664 (1991).
  \bibitem{TELEPORTATION}
  C. H. Bennett, et al., Phys.
Rev. Lett. {\bf 70}, 1895--1898 (1993).
\bibitem{GILL}
  R. D. Gill, G. Weihs, A. Zeilinger, and M. \.Zukowski,  Proc. Nat. Acad. Sci. {\bf 99}, 14632--14635 (2002).
\bibitem{ROTATIONAL}
K. Nagata, W. Laskowski , M. Wie\'sniak, M. \.Zukowski, Phys. Rev. Lett. {\bf 93}, 230403 (2004).
\bibitem{ENTANGLEMENT}
P. Badziag, C. Brukner, W. Laskowski, T. Paterek, M \.Zukowski, Phys. Rev. Lett, {\bf 100}, 140403 (2008).
\bibitem{HORODECKI}
M. Horodecki, P. Horodecki, and R. Horodecki,
Phys. Lett. A {\bf 223}, 1--6 (1996); 
 A. Peres,
Phys. Rev. Lett. {\bf 77}, 1413-1415 (1996); R. Horodecki, P. Horodecki, M. Horodecki, and K. Horodecki,
e-print  arXiv:quant-ph/0702225. 
\bibitem{COMPLEXITY-1}\v{C}. Brukner, M. \.Zukowski, A. Zeilinger, Phys. Rev. Lett. {\bf
89}, 197901 (2002);
\v{C}. Brukner {\it et al.}, Phys. Rev. Lett. {\bf 92}, 127901
(2004).
\bibitem{COMPLEXITY-2}
P. Trojek, Ch. Schmid, M. Bourennane, C.
 Brukner, M. \.Zukowski, H. Weinfurter, Phys. Rev. A {\bf 72}, 050305R (2005).
\bibitem{ZUKOWSKI-1993}
M.~\.Zukowski,  Phys. Lett . A {\bf 177}, 290--297 (1993).
\bibitem{WEIHS}
G.~Weihs, T.~Jennewein, C.~Simon, H.~Weinfurter, and
A.~Zeilinger, Phys. Rev. Lett. {\bf 81}, 5039 (1998).
\bibitem{XXX}
 D. N. Matsukevich, P. Maunz, D. L. Moehring, S. Olmschenk, C. Monroe, Phys. Rev. Lett. {\bf 100}, 150404 (2008). 


\bibitem{GISIN}
N.~Gisin, 
Phys. Lett. A, {\bf 154}, 201--203 (1991). 

\bibitem{NATURE} 
S.~Groeblacher, T.~Paterek, R.~Kaltenbaek R, C.~Brukner, M.~\.Zukowski, M.~Aspelmeyer, A.~Zeilinger,  Nature {\bf 446},
871--875 (2007).
\bibitem{WERNER}
 R.F. Werner, 
Phys. Rev. A    {\bf 40}, 4277--4280 (1989). 

\bibitem{YAO}
A. C.-C. Yao, in {\it Proceedings of the 11th Annual ACM Symposium
on Theory of Computing}, pp. 209--213, 1979;
E. Kushilevitz and N. Nisan, {\it Communication complexity}
Cambridge University Press, Cambridge 1997.
\bibitem{BURN}
H. Buhrman, {\it et. al}, e-print quant-ph/9705033; H. Buhrman
{\it et. al}, Phys. Rev. A {\bf 60}, 2737--2742 (1999);
R. Cleve and H. Buhrman, Phys. Rev. A {\bf 56}, 1201 (1997); L. K.
Grover, e-print quant-ph/9704012; G. Brassard, e-print
quant-ph/0101005.
\bibitem{GALVAO}
E. F. Galv\~{a}o, Phys. Rev. A. {\bf 65}, 012318 (2001);
\bibitem{DISTILL}
 C. H. Bennett, G. Brassard, S. Popescu, B. Schumacher,
J. A. Smolin, and W. K. Wootters, Phys. Rev. Lett. {\bf 76}, 722--725
(1996). - Erratum: Phys. Rev. Lett. {\bf 78}, 2031 (1997).
\bibitem{PATEREK}
T. Paterek, C. Brukner , M. Zukowski,  Int. J. Quant. Inf. {\bf 1}, 519--528 (2003).   
\end{thebibliography}
\end{document}